\documentstyle[pra,aps,epsfig,float]{revtex}
\begin{document}
\draft
\wideabs{
\title{Implementing unitary operators in quantum computation}
\author{Jaehyun~Kim, Jae-Seung~Lee, and Soonchil~Lee}
\address{Department of Physics, Korea Advanced Institute of Science %
and Technology, Taejon 305-701, Korea}
\date{\today}
\maketitle
\begin{abstract}
We present a general method which expresses a unitary operator by the product of
operators allowed by the Hamiltonian of spin-1/2 systems. In this method, the
generator of an operator is found first, and then the generator is expanded by the
base operators of the product operator formalism. Finally, the base operators
disallowed by the Hamiltonian, including more than two-body interaction operators,
are replaced by allowed ones by the axes transformation and coupling order
reduction technique. This method directly provides pulse sequences for the
nuclear magnetic resonance quantum computer, and can be generally applied to other
systems.
\end{abstract}

}

\section{Introduction}
In 1973, Bennett~\cite{bennett73} proposed a reversible Turing machine that is as
efficient as an irreversible one, and this led to the idea of using a quantum system
as a computer because the time evolution of a quantum system is reversible.
Feynman~\cite{feynman82} introduced the concept of a quantum computer, and its
theoretical model was given by Deutsch~\cite{deutsch85}. On the other hand, Fredkin
and Toffoli~\cite{fandt} proved that an arbitrary computation can be performed by a
reversible Turing machine by showing that {\sc and}, {\sc or}, and {\sc not} gates
can be generated by reversible 3-bit gates among which a Toffoli
gate~\cite{toffoli} is most frequently used nowadays. In quantum computation, a
three-bit gate cannot be implemented directly because it requires a simultaneous
interaction of three particles. Thus there have been efforts to find two-bit
universal
gates~\cite{deutsch89,divincenzo,deutsch95,barenco1,lloyd1,barenco2,sleator}.
In particular, Barenco {\it et al.} showed that a combination of two-bit c-{\sc not}
gates and one-bit gates can replace a Toffoli gate, and proposed a method to make
general $n$-bit controlled gates~\cite{barenco}. Therefore, it is proved that an
arbitrary computation can be performed by a quantum computer, and the
implementation of these universal gates became the basic requirement for any
quantum system to be a quantum computer.

However, the proof that an arbitrary computation can be done by a quantum computer
does not necessarily mean that we know a general implementation procedure. If n
unitary operator $U$, equivalent to a combination of gates, is related to the
Hamiltonian ${\cal H}$ of a certain quantum system by $U=\exp (-\imath {\cal
H}t/\hbar )$, it can be realized by the time evolution of the system during time
$t$. But there are only a few operations that can be implemented in this way by the
limited Hamiltonians of nature. Therefore, it is very necessary to find a general
method to implement an arbitrary operation using only the given Hamiltonians.
Feynman proposed a way to construct an artificial Hamiltonian when $U$ is given by
$U=U_k\cdots U_3U_2U_1$ and all ${\cal H}_i$'s corresponding to $U_i$'s exist in
nature~\cite{feynman96}, but it is impractical to construct artificial
Hamiltonians. It will be more practical to partially control a Hamiltonian by
turning perturbations ``on'' and ``off'' if $U$ can be expressed as a product of
operators corresponding to the perturbation terms. Whether Feynman's artificial
Hamiltonians or switchable perturbations are used, an operator of interest must be
expressed as a product of the operators allowed by Hamiltonians. This is equivalent
to finding the combination of universal gates, or a quantum network, and generally
is a very difficult problem having several solutions.

In this work, we propose a general method of expressing a unitary operator as a
product of the operators allowed by the Hamiltonian of the spin-$\frac12$ systems
including the nuclear magnetic resonance (NMR) quantum computer. This method makes
use of the fact that a unitary operator $U$ is always given by $U=\exp[-\imath
G]$, where $G$ is a Hermitian operator. Once the generator of an operator, $G$, is
found, it is expanded by suitable base operators. Then $U$ is expressed as a
product of operators having only one base operator as a generator and, finally,
each operator in the product is replaced by the allowed ones. Compared to a
previous report~\cite{tucci}, only the operators of physical variables were used in
each transformation procedure. This helps to understand the physical meaning of
operations done by a quantum computer.

\section{decomposition}
The first step of implementation is to find the generator of a given operator.
Since the only way to implement an operator is to use the time evolution of a state
under a suitable Hamiltonian, a generator, which is the product of Hamiltonian and
time, gives the physical information necessary for implementation. A unitary operator
is represented by a normal matrix and always diagonalized by unitary
transformation. The matrix $T$ that diagonalizes $U$ also diagonalizes $G$ as
\begin{equation}
U' = TUT^\dagger = e^{-\imath TGT^\dagger} = e^{-\imath G'},
\end{equation}
where $U'$ and $G'$ are diagonalized matrices of $U$ and $G$, respectively.
Once the operator and its generator become diagonal, $G'$ is easily obtained from
\begin{equation}
U'_{kk} = e^{-\imath G'_{kk}}
\label{eq-map}
\end{equation}
and $G$ is obtained by inverse transformation $G = T^\dagger G'T$. Since $G$ is
Hermitian, $G'_{kk}$, the eigenvalues of $G$, are real and $U'_{kk}$ are complex
with absolute value of unity. It is worthwhile to note that the mapping from
$U'_{kk}$ to $G'_{kk}$ is not unique.

To relate the generator $G$ with Hamiltonians, consider the following operators of
the product operator formalism for $N$ spin-$\frac12$
particles~\cite{sorensen,ernst,somaroo}:
\begin{equation}
B_s = 2^{(q-1)}(I_{\alpha_1}\otimes I_{\alpha_2}\otimes\cdots
      \otimes I_{\alpha_N}),
\label{eq-basis}
\end{equation}
where $s=\{\alpha_1, \alpha_2,...,\alpha_N\}$ and $\alpha_i$
is $0$, $x$, $y$, or $z$. $I_0$ is $E$, i.e., a $2\times 2$ unity
matrix, $I_{\alpha_i}$ is a spin angular momentum operator for
$\alpha_i \neq 0$, and $q$ is the number of nonzero $\alpha_i$'s.
For example, $\{B_s\}$ for $N=2$ is given by
\begin{equation}
\begin{array}{rl}
q=0;& E/2 \\
q=1;& I_{1x}, I_{1y}, I_{1z}, I_{2x}, I_{2y}, I_{2z} \\
q=2;& 2I_{1x}I_{2x}, 2I_{1x}I_{2y}, 2I_{1x}I_{2z},...,
\end{array}
\label{eq-prod}
\end{equation}
which are 16 Dirac matrices except the factor of $\frac12$. In Eq.~(\ref{eq-prod}),
unity matrices are not shown and spin indices are added for convenience. $\{B_s\}$,
consisting of $4^N$ elements, makes a complete set and, therefore, an arbitrary
$2^N\times 2^N$ matrix can be expanded by the linear combination of $B_s$'s. Since
$G$ and $B_s$'s are Hermitian, coefficients of the linear expansion are real
numbers and obtained by applying the inner product of $G$ and $B_s$'s.

A unitary operator is now expressed as $U = \exp(- \imath \sum_s b_s B_s )$, of
which the generator is related to physical observables. In general, there exists no
Hamiltonian that corresponds to a linear combination of $B_s$'s. Therefore, our
next step is to express $U$ as a product of {\it single operators}, which have only
one $B_s$ as a generator like $\exp[-\imath b_s B_s]$. Sometimes this decomposition
is the most difficult step, and it has not yet been proven whether the decomposition is
generally possible even for spin operators. Fortunately, many useful gates can be
easily decomposed by using the commutation relations of $B_s$'s. $B_s$'s are either
commuting or anticommuting with each other. If $G$ is expanded with only commuting
$B_s$'s, $U$ can be easily represented by a product of single operators as
\begin{equation}
U=\exp [-\imath\sum_s b_sB_s] \rightarrow \prod_s \exp [-\imath b_sB_s].
\end{equation}
A swap gate and an $f$-controlled phase shift gate used in
Grover's search algorithm belong to this case.

Even though a generator has non-commuting $B_s$'s, there are cases
where decomposition is straightforward. Suppose two base
operators $B_{s1}$ and $B_{s2}$ satisfy the relation ($\hbar = 1$):
\begin{equation}
[B_{s1},B_{s2}]=\imath B_{s3};
\label{eq-commrel}
\end{equation}
then $B_{s3}$ also belongs to $\{B_s\}$. This commutation relation makes the three
operators $B_{s1}$, $B_{s2}$, and $B_{s3}$ transform like Cartesian coordinates
under rotation, meaning that
\begin{equation}
\exp[-\imath\phi B_{s3}]B_{s1}(\exp[-\imath\phi B_{s3}])^\dagger
 = B_{s1}\cos\phi + B_{s2}\sin\phi
\end{equation}
for cyclic permutations of $s1$, $s2$, and $s3$. If a generator has only these
operators, it can be decomposed using Euler rotations. For example,
$\exp[-\imath\phi(B_{s1}+B_{s2})]$ is understood to be the rotation with the angle
of $\sqrt2\phi$ about the axis $45^\circ$ off the ``$B_{s1}$ axis'' on the plane of
$B_{s1}$ and $B_{s2}$ axes. Therefore, this operation is equivalent to the
successive rotations about $B_{s1}$ and $B_{s3}$ axes as follows:
\begin{equation}
e^{-\imath\phi(B_{s1}+B_{s2})} = e^{-\imath\frac{\pi}{4}B_{s3}}
 e^{-\imath\sqrt2\phi B_{s1}}e^{\imath\frac{\pi}{4}B_{s3}}.
\label{eq-euler}
\end{equation}
This decomposition technique by Euler rotations is also applicable when an operator
has a generator in the factorized form as follows:
\begin{equation}
U = \exp\left[-\imath \prod^N_{i=1} \left(\sum_{\alpha_i}
\phi_{i\alpha_{i}}I_{i\alpha_{i}}\right)\right],
\label{eq-facgen}
\end{equation}
where $\phi_{i\alpha_{i}}$ are real numbers. Since $I_{1x}$, $I_{1y}$, and $I_{1z}$
satisfy the commutation relation in Eq.~(\ref{eq-commrel}), and commute with any
other spin operators with $i\neq 1$, spin-1 components are decomposed as
\begin{equation}
U_1\exp\left[-\imath(\phi_{10}E+\phi_1I_{1\alpha_{1}})
\prod^N_{i=2} \left(\sum_{\alpha_{i}}
\phi_{i\alpha_{i}}I_{i\alpha_{i}}\right)\right]U^\dagger_1,
\label{eq-fac2}
\end{equation}
where $U_1$ is the product of the single operators of which the generators have
only spin-1 components, corresponding to Euler rotations. Repeated applications of
this process to successive spins give
\begin{equation}
U= U_N\cdots U_1 e^{-\imath G} U^\dagger_1\cdots U^\dagger_N,
\end{equation}
where
\begin{equation}
G = \prod^N_{i=1}(\phi_{i0}E+\phi_i I_{i\alpha_{i}}).
\label{eq-fac3}
\end{equation}
Then decomposition is finished because all terms in Eq.~(\ref{eq-fac3}) commute with
each other. All the controlled gates belong to this case. If none of the above
methods are applicable, $U$ can be approximately expanded as a product of single
operators to any desired accuracy~\cite{sorn}.

\section{reduction}
Although $B_s$ is a product of spin operators that are physical quantities, not all
$B_s$'s exist in Hamiltonians that nature allows. The next step of implementation
is to replace disallowed single operators in the product by allowed ones. The
Hamiltonian of a spin-$\frac12$ system used for implementation of a quantum
computer allows only the following single operators in general.
\begin{equation}
\begin{array}{l}
R_{i\alpha}(\phi)=e^{-\imath\phi I_{i\alpha}}, \\
J_{ij\alpha}(\phi) = e^{-\imath\phi 2I_{i\alpha}I_{j\alpha}}.
\end{array}
\end{equation}
The first term is a rotation operator that rotates spin $i$ about the $\alpha$ axis by
the angle of $\phi$, and the second one is a spin-spin interaction operator between
spins $i$ and $j$. The angle $\phi$ in the second term is proportional to the
spin-spin coupling constant and evolution time, but we denote it as a rotation
angle because the effect of spin-spin interaction can be understood as a rotation
of one spin due to the magnetic field of the other. Before going further, we assume
the following more restricted set of operators as allowed ones in this study:
\begin{equation}
\begin{array}{c}
R_{i\alpha}(\phi)=e^{-\imath\phi I_{i\alpha}}~~(\alpha = x~{\rm or}~y), \\
J_{ij}(\phi) = J_{ijz}(\phi) = e^{-\imath\phi 2I_{iz}I_{jz}}.
\end{array}
\label{eq-bop}
\end{equation}
In this set, only $x$ and $y$ axes are used for single-spin rotations and a
spin-spin interaction is limited to the Ising type. Needless to say, the greater
the number of single operators allowed, the easier it is to implement an algorithm.
However, Eq.~(\ref{eq-bop}) is a sufficient set to realize any unitary operators as
shown below, and in fact these are the only operators allowed by an NMR quantum
computer. Single-spin rotations are implemented by selective rf pulses and
spin-spin interactions by Hamiltonian evolution with intermediate refocusing
pulses~\cite{linden2}. These two rotation operators can generate any single bit
operation and the interaction operator can make a $c$-{\sc not} gate in combination
with rotation operators~\cite{linden2,jones3}. Therefore, these three operators
consist the minimum set to implement universal gates.

Now, we are to show that the minimum set in Eq.~(\ref{eq-bop}) can generate all the
other single operators. First, the-single bit operation excluded in
Eq.~(\ref{eq-bop}), $R_{iz}(\phi)$, can be transformed from $R_{ix}(\phi)$ as
\begin{equation}
R_{iz}(\phi) = R_{iy}\left(-\frac{\pi}{2}\right) R_{ix}(\phi)
  R_{iy}\left(\frac{\pi}{2}\right).
\end{equation}
This is the composite pulse technique well-known in the NMR
experiments~\cite{ernst}. Any rotation about one axis can be replaced by the
composite of rotations about the other two. All the second-order operators, where
the $n$th-order operator means the single operator that has a generator $B_s$ with
$q=n$, can be transformed into the Ising-type operator in Eq.~(\ref{eq-bop}) by
this technique~\cite{havel,price1}. For example, $\exp[-\imath\phi 2I_{ix}I_{jz}]$
is transformed as
\begin{equation}
\exp[-\imath\phi 2I_{ix}I_{jz}]
  = R_{iy}\left(\frac{\pi}{2}\right)\exp[-\imath\phi 2I_{iz}I_{jz}] R_{iy}\left(-\frac{\pi}{2}\right).
\end{equation}
In the same way, any $n$th-order operator can be transformed into the product of
single operators and the $n$th-order coupling operator that is defined as the
$n$th-order operator with all $\alpha_i=z$.

After all the spin coordinates are changed to $z$ using this technique, the
operators with more than two-body interaction can be reduced to an Ising-type
two-body interaction operator as discussed below. The key idea of the coupling
order reduction is that the $n$th-order coupling operator can be thought as the
$(n-1)$th-order one controlled by one spin state. For example, the third-order
coupling operator~\cite{tseng}, $\exp[-\imath\phi 4I_{iz}I_{jz}I_{kz}]$ is
represented by
\begin{equation}
\begin{array}{l}
\exp[-\imath\phi 4I_{iz}I_{jz}I_{kz}] \\
~~~= \exp\left[ -\imath \phi
          \left( \begin{array}{cc} 2(I_z\otimes I_z) & 0 \\
          0 & -2(I_z\otimes I_z) \end{array} \right) \right] \nonumber \\
~~~= \left( \begin{array}{cc} \exp[-\imath\phi 2(I_z\otimes I_z)] & 0 \\
          0 & \exp[\imath\phi 2(I_z\otimes I_z)] \end{array} \right) \\
~~~= \left( \begin{array}{cc} J_{jk}(\phi) & 0 \\
          0 & J_{jk}(-\phi) \end{array} \right)
\end{array}
\label{eq-3rd}
\end{equation}
in the subspace of spin $i$. The final form of Eq.~(\ref{eq-3rd}) implies that the
third-order coupling operator can be understood as a second-order one with coupling
between spin $j$ and $k$, but its rotation direction depends on the state of spin
$i$. We note that if one spin is flipped during the evolution of the spin-spin
interaction, then the sign of the interaction changes and this has the effect of
time reversal. This means that the rotation direction changes~\cite{ernst,linden2}
and, therefore, we can implement Eq.~(\ref{eq-3rd}) with the second-order coupling
operator by flipping spin $j$ or $k$ depending on the state of spin $i$. It is a
well-known $c$-{\sc not} ({\sc xor}) gate that flips one spin depending on the
state of the other spin. A $c$-{\sc not} gate is given by
\begin{equation}
U_{\text{c-NOT}} = R_{iz}\left(\frac{\pi}{2}\right)
   R_{jx}\left(\frac{\pi}{2}\right) R_{jy}\left(\frac{\pi}{2}\right)
   J_{ij}\left(-\frac{\pi}{2}\right) R_{jy}\left(-\frac{\pi}{2}\right)
\label{eq-xor}
\end{equation}
up to an overall phase, and this is the product of allowed operators in
Eq.~(\ref{eq-bop}).

\begin{figure}[t]
\centering
\epsfig{file=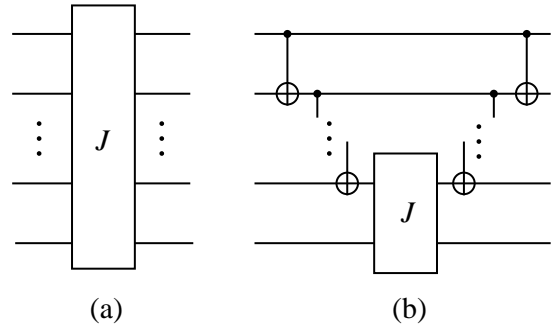, width=8cm}
\caption{\narrowtext Quantum network for the $n$th-order coupling operator (a) %
and its equivalent network consisting of allowed operators (b).} \label{fig-1}
\end{figure}

In the same way, an $n$th-order coupling operator can be reduced to an
$(n-1)$th-order one by conditionally flipping odd number of spins except the spin
$i$. Repeated applications of this process obviously reduce an $n$th-order coupling
operator to a second-order one. Fig.~\ref{fig-1} shows the quantum networks of the
$n$th-order coupling operator and its equivalent combination of the allowed
operators. In Fig.~\ref{fig-1}(b), the $c$-{\sc not} gates after the second-order
coupling operator are inserted to flip spins to their original states. Instead of
the $c$-{\sc not} gates before and after the second-order coupling operator, a
pseudo $c$-{\sc not} gate
$U_{ij}=R_{jx}(\frac{\pi}{2})J_{ij}(\frac{\pi}{2})R_{jy}(\frac{\pi}{2})$ and
$U^\dagger_{ij}$ can be used, respectively.

As an example, we apply this general implementation procedure to a Toffoli
gate~\cite{price2}. The generator of a Toffoli gate obtained after the processes of
diagonalization and inverse unitary transformation is expanded by base operators as
\begin{equation}
\begin{array}{l}
G = \pi(\mbox{}-\frac18 E + \frac14 I_{1z} + \frac14 I_{2z}
 - \frac14 2I_{1z}I_{2z} + \frac14 I_{3x} \\
~~~~~~~ \mbox{}- \frac14 2I_{1z}I_{3x} - \frac14 2I_{2z}I_{3x}
 + \frac14 4I_{1z}I_{2z}I_{3x}).
\end{array}
\label{eq-tof}
\end{equation}
Since all terms in this generator commute with each other, the corresponding
operator is easily expressed by the product of single operators. After replacing
disallowed operators, $I_{1z}I_{3x}$ and $I_{1z}I_{2z}I_{3x}$ in this case, by
allowed ones by axes transformation and coupling order reduction, the gate is
finally expressed as
\begin{equation}
\begin{array}{l}
R_{1z}(\frac{\pi}{4})R_{2z}(\frac{\pi}{4})J_{12}(-\frac{\pi}{4})
R_{3x}(\frac{\pi}{4})R_{3y}(\frac{\pi}{2})J_{13}(-\frac{\pi}{4}) \\ \times
J_{23}(-\frac{\pi}{4})U_{12}J_{23}(\frac{\pi}{4})U^\dagger_{12}
R_{3y}(-\frac{\pi}{2})
\end{array}
\end{equation}
up to an overall phase.

\section{conclusion}
Our method is applicable to any quantum computers that use spin-$\frac12$ states as
qubits, because the operators of Eq.~(\ref{eq-bop}) make the minimum set required
to those quantum computers. This method can be generalized for other quantum
computer systems that should provide a complete set of operators similar to those
of Eq.~(\ref{eq-bop}). Since our method uses generators that are closely related to
Hamiltonians, this method helps us to see the physical meaning of an operation.
Operators with generators disallowed by Hamiltonians are replaced by allowed ones
using axes transformation and order reduction techniques. Therefore, it is possible
to simulate an Hamiltonian that does not exist in nature using this method,
including more than two-body interactions.

This method does not necessarily give either optimal or unique solution to
implementation. As the number of spins increases, the number of base operators
grows exponentially, and a generator could have too many terms. Therefore, it is
impractical to apply this method to a system with many spins, but our method still
provides a good guide to implement an operator of interest.

\end{document}